\documentclass[aps,preprintnumbers,amsmath,nofootinbib]{revtex4}
\usepackage{amssymb}
\usepackage{graphicx}
\usepackage{dcolumn}
\usepackage{bm}

\begin{document}
\title
{Dependence of soft pion jet properties in the space of relative
four-dimensional velocities on initial energy}

\author{V.A. Okorokov} \email{VAOkorokov@mephi.ru; okorokov@bnl.gov}
\affiliation{National Research Nuclear University "MEPhI",
Kashirskoe Shosse 31, 115409 Moscow, Russia}

\date{\today}

\begin{abstract}
In the paper experimental results obtained by studying of
collective and fractal properties of soft pion jets in the space
of relative four-dimensional velocities in intermediate energy
domain 2 -- 20 GeV are presented. Fractional values of cluster
dimension are indicated on manifestation of fractal-like
properties by pion jets. The changes of the mean square of the
distance between secondary particles and jet axis, the mean
kinetic energy of particles in jets and the cluster dimension with
increasing of collision energy agree with the hypothesis of
manifestation of quark degrees of freedom in processes of pion jet
production at intermediate energies. For the first time the
quantitative estimations are obtained for the low boundary energy
at which quark degrees of freedom start to display itself in
production of soft pion jets experimentally. The value of
estimation for this parameter derived with taking into account of
all used collective parameters is $(2.8 \pm 0.6)$ GeV.

\textbf{PACS} 13.75.-n,
13.85.-t
\end{abstract}

\maketitle

\section{Introduction}\label{intro}
At the present time, the decision of the problem of confinement
and study of transition from meson-baryon degrees of freedom to
quark-gluon ones is one of the most important (and, at the same
time, most difficult) task of world research programm in the field
of strong interactions. The nature of confinement of color degrees
of freedom (quark and gluons) and, correspondingly, the possible
phase transitions in strongly interacting matter is not clear in
full so far. Observation of hadron jets at high energies is one of
the most important and evident experimental manifestation of
quark-gluon degrees of freedom. At present, an open question is
where is the low boundary on energy starting with which the color
degrees of freedom should be taken into account for description of
processes of multiparticle production. Obviously, the jet
structure of events displays itself more clear at high initial
energies ($\sqrt{s}$) than at intermediate ones. But in despite of
this feature it seems that the application of collective
characteristics of multiparticle final state can be useful in
collision energy domain $\sqrt{s}=2-20$ GeV both for study of
transition from the predominance of meson-baryon degrees of
freedom to quark-gluon ones and for qualitative estimation of low
boundary (for initial energy) experimental manifestation of quark
degrees of freedom in soft jet production. In various fields of
physics the onset of manifestation of new degrees of freedom and
transition processes is accompanied by the presence of self-affine
and fractal properties in collective effects. Therefore, the study
of collective and geometric (fractal-like) properties, in
particular, of soft pion jets at intermediate energies can give
the new important information about hadronization mechanisms,
non-perturbative physics and transition to manifestation of quark
degrees of freedom in collective phenomena.

The paper is organized as follows. In Sec.\,2, definitions of
collective variables are described. The Sec.\,3 devotes discussion
of energy dependence of the properties of soft pion jets in the
space of relative four-dimensional velocities, estimations of low
boundary energy for quark degrees of freedom experimental
manifestation in pion jet production based on empirical
approximations of experimental data. Some final remarks and
conclusions are presented in Sec.\,4.

\section{Method and variables}\label{sec:2}
Traditional collective characteristics used for study of event
shape \cite{Okorokov-IJMPA-27-1250037-2012} are not
relativistically invariant. This introduces some additional
kinematic uncertainties, for example, in choice of center-of-mass
system for reactions with atomic nuclei. A relativistically
invariant method was proposed in
\cite{Baldin-DoklANUSSR-222-1064-1975} for investigation of
collective effects in multiparticle production processes of the
following type: $\mbox{b} + \mbox{t} \to 1+2+...$, where
$\mbox{t\,/\,b}$ -- target\,/\,beam particle. This method based on
the using of new variables allows to obtain relativistically
invariant description of excited strongly interacting matter in
wide range of initial energies and for different reaction types.
Numerous results of study of jet production in soft hadron
collisions shown that there are two jets emitted in opposite
directions in forward and backward hemispheres in these processes
(see, for example, \cite{Okorokov-PhD-1996}). Special features of
relativistically invariant method in this case were considered in
detail elsewhere
\cite{Okorokov-YaF-57-2225-1994,Okorokov-YaF-62-1787-1999}. It
should be emphasized the method under study is most rigorously
justified and applied successfully in the case of two-jet event
shape which corresponds to the geometry of final state in the
domain of intermediate energies. In the case, the separation of
secondary particles on fragmentation regions can be made with the
aid of the relativistically invariant variables
$X^{i}_{\mbox{\footnotesize{b}}}$ and
$X^{i}_{\mbox{\footnotesize{t}}}$, which are defined as following
\cite{Baldin-YaF-44-1209-1986,Baldin-PreprintJINR-E1-87-142-1987}
$ X^{i}_{\alpha}=[m_{i}(U_{i}U_{\beta})] /
[m_{\alpha}(U_{\alpha}U_{\beta})];~ \alpha,
\beta=\mbox{t,b};~\alpha \ne \beta. $ Here $m_{i}$ is the mass of
$i^{\,\mbox{\footnotesize{th}}}$ secondary particle,
$m_{\mbox{\footnotesize{t\,/\,b}}}$ -- mass of target\,/\,beam
particle, $P_{i},~U_{i}=P_{i}/m_{i},~i=\mbox{b, t},1,2,...$ --
four-momenta and four-velocities. The variables
$X^{i}_{\mbox{\footnotesize{t/b}}}$ characterize the fraction of
four-momentum of initial particles ($P_{i},~i=\mbox{t\,/\,b}$)
carried away by secondary hadrons in the target\,/\,beam
fragmentation region respectively. Fragmentation regions in the
plane $(X^{i}_{\mbox{\footnotesize{b}}},
X^{i}_{\mbox{\footnotesize{t}}})$ are defined by the following
sets of conditions
\cite{Baldin-YaF-44-1209-1986,Baldin-PreprintJINR-E1-87-142-1987}:
\begin{center}
target fragmentation -- $\left \{ \begin{array}{lcl}
X^{i}_{\mbox{\footnotesize{t}}}&\geq& \tilde{X},
\vspace{0.1cm} \\
X^{i}_{\mbox{\footnotesize{b}}}&\leq& \tilde{X}; \\
\end{array}
\right. $~~~~~ \hspace{1.0cm}beam fragmentation -- $\left \{
\begin{array}{lcl} X^{i}_{\mbox{\footnotesize{t}}}&\leq&
\tilde{X},
\vspace{0.1cm} \\
X^{i}_{\mbox{\footnotesize{b}}}&\geq& \tilde{X}. \\
\end{array}
\right. $
\end{center}
Here $\tilde{X}=0.1-0.2$ is some boundary value which is
determined empirically. The basic quantities which the probability
distributions (cross sections) depend upon are non-dimension
positive relativistic invariant quantities
$b_{ik}=-(U_{i}-U_{k})^{2}$, where $i, k=\mbox{b, t},1,2,...$
\cite{Baldin-DoklANUSSR-222-1064-1975}. As seen the observables
$b_{ik}$ mean the squares of relative distances in the
four-velocities space. The comparison of this method for
distinguishing of some particle groups in the space of
four-dimensional velocities with other present non-invariant
(traditional) methods allows to name these separate groups as jets
\cite{Baldin-YaF-44-1209-1986}. The jet axis in the space under
study for certain fragmentation region, $V$, is defined by unit
vector of jet four-velocity: $V = U_{\mbox{\footnotesize{J}}} /
|U_{\mbox{\footnotesize{J}}}|$, where
$U_{\mbox{\footnotesize{J}}}=\sum_{k=1}^{N}U_{k}$, $N$ is the
number of particles in considered fragmentation region which are
satisfied all cuts and involved in the analysis. One of the most
important observables of this method is the square of the distance
of the $k^{\,\mbox{\footnotesize{th}}}$ particle from the jet axis
\cite{Baldin-DoklANUSSR-222-1064-1975}
\begin{equation}
b_{k}=-\left(V-U_{k}\right)^{2},~k=\mbox{t,b},1,2,... \label{eq:1}
\end{equation}
The study of properties of soft pion jets is fulfilled with the
aim of invariant function $F(b_{k})$ introduced for nucleon
clusters in
\cite{Baldin-Lektcii-43-P1-87-912-1987,Baldin-FortschrPhys-38-1990-261}.
For pion jets this function is re-defined as following
\cite{Okorokov-YaF-62-1787-1999} $F(b_{k}) = (\lambda / N)dN /
db_{k},~\lambda \equiv 4 / [m_{\pi}^{2}b_{k}\sqrt{1+4/b_{k}}]$.
The investigations of $F(b_{k})$ both for nucleon clusters
\cite{Baldin-Lektcii-43-P1-87-912-1987,Baldin-FortschrPhys-38-1990-261}
and for pion jets
\cite{Okorokov-YaF-62-1787-1999,Baldin-ECHAYA-29-577-1998}
demonstrated that the $F(b_{k})$ can be approximated by exponent
function $F^{\alpha}(b_{k}) =
\sum_{i=1}^{m}p_{i}^{\alpha}\exp(-b_{k}/\langle
b_{k}\rangle_{i}^{\alpha})$ quite reasonably, where
$\alpha=\mbox{t\,/\,b}, m=1$ or $m=2$ depending on certain
fragmentation region, reaction type and $\sqrt{s}$. The mean
kinetic energy of particles in the jet in its rest frame (it is
referred to as "temperature"), $\langle T_{k}\rangle$, is an
important characteristic of particles inside jet which is
determined on the basis of above approximation of invariant
$F(b_{k})$ distributions as following
\begin{equation}
\langle T_{k}\rangle^{\alpha}_{i} = m_{\pi}\langle
b_{k}\rangle^{\alpha}_{i} / 2. \label{eq:2}
\end{equation}
In \cite{Okorokov-NSMEPhI-218-2000} it was proposed to study
geometric properties of pion jets in the space of relative
four-velocities with the aid of the cluster dimension $D$, which
is determined by the relation between the number of particles in
the jet, $n_{\mbox{\scriptsize{J}}}(b_{k})=\displaystyle
\int_{0}^{b_{k}} db'_{k}\, dN/db'_{k}$, and its radius in the
space under discussion. Taking into account that (\ref{eq:1}) is
the square of distance in the space of relative four-velocities
this relation is
$n_{\mbox{\scriptsize{J}}}(b_{k})=(ab_{k})^{D/\,2}$ at
$n_{\mbox{\scriptsize{J}}} \to \infty$, which is in reasonable
agreement with corresponding experimental dependencies
\cite{Okorokov-NSMEPhI-218-2000,Okorokov-YaF-73-2016-2010}. Thus
the cluster dimension of pion jet in the space of relative
four-velocities can be derived as following
\begin{equation}
D = 2\ln [n_{\mbox{\scriptsize{J}}}(b_{k})] / \ln (ab_{k}).
\label{eq:3}
\end{equation}
One need to note that the parameter (\ref{eq:3}) is an integer and
coincides with the topology dimension ($D{\mbox{\scriptsize{T}}}$)
of space under consideration if particle distribution is
homogeneous in this space. But $D$ can be non-integer and $D \ne
D{\mbox{\scriptsize{T}}}$ for particle distribution with highly
irregular and complex structure. Thus non-integer value of cluster
dimension can be considered as characteristic signature of
manifestation of fractal-like properties \cite{Fomenko-book-1998}.
For most complex distributions the multi-fractal structure can be
appeared and cluster dimension be a function of jet radius for
such case: $D=D(b_{k})$. Thus the (\ref{eq:3}) is the qualitative
parameter reflected the features of particle distribution in phase
space.

It should be emphasized that in simplest case of lowest order of
renormalization group equation (RGE) the following relation can be
derived $\alpha_{S} \propto 1/\ln b_{ik}$, where
$i=\mbox{t\,/\,b}$, $\alpha_{S}$ is the renormalized strong
coupling constant
\cite{Baldin-Lektcii-43-P1-87-912-1987,Baldin-FortschrPhys-38-1990-261,Baldin-NPA-434-695c-1985}.
It was shown
\cite{Baldin-Lektcii-43-P1-87-912-1987,Baldin-FortschrPhys-38-1990-261,Baldin-NPA-434-695c-1985}
that the $b_{ik}$ values can be used to classify relativistic
nuclear interactions: the domain $b_{ik} \sim 10^{-2}$ corresponds
to the interaction of nuclei as weakly bound systems of nucleons
(domain of classic nuclear physics), the region $b_{ik} > 1$
corresponds to the interaction of hadrons as weakly bound quark
systems (domain of QCD). Within the approach under study the range
of values $10^{-2} \lesssim b_{ik} \sim 1$ corresponds to the
transition from the domain of dominance of meson-baryon degrees of
freedom to the region where the internal structure of colliding
particles and, as consequence, quark-gluon degrees of freedom
become essential in the processes of secondary particle jet
production\footnote{One need to note that the boundary values for
$b_{k}$ indicated above are qualitative phenomenological
estimations.}. It should be emphasized that the estimations for
boundaries of various dynamic domains in the space of relative
four-velocities indicated above seem valid for $b_{k}$, i.e. when
the jet axis is corresponds to the "reference"
$i^{\,\mbox{\footnotesize{th}}}$ particle. But the possible exact
relation between $b_{k}$ and $\alpha_{S}$ is required the
additional rigorous substantiation and careful derivation. Here
one can note only that jets consist of particles with close masses
(pions) in the present study and there is the relation $\langle
b_{k}\rangle = 2(\langle M_{\mbox{\scriptsize{J}}}\rangle /
\langle n_{\mbox{\scriptsize{J}}}\rangle m_{h}-1)$ between the
mean square of jet size in the space of relative four-velocities
and mean effective mass ($M_{\mbox{\scriptsize{J}}}$) of jets of
secondary hadrons with identical masses $m_{h}$
\cite{Grishin-Book-1988,Okorokov-ISHEPP-154-2006}. Here $\langle
n_{\mbox{\scriptsize{J}}}\rangle$ is the mean multiplicity of
particles inside jet, the averaging is taken over particles in
event and over events in sample in the l.h.s.; over event ensemble
in the r.h.s. In collider experiments at high energies the jet
transverse momentum
($p_{\mbox{\scriptsize{J}}}^{\mbox{\scriptsize{T}}}$)
\cite{D0-PLB-718-56-2012} or average transverse momentum of the
two jets leading in
$p_{\mbox{\scriptsize{J}}}^{\mbox{\scriptsize{T}}}$ ($\langle
p_{\mbox{\scriptsize{J1,2}}}^{\mbox{\scriptsize{T}}}\rangle$)
\cite{CMS-arXiv-1304.7498-2013} are considered as scale of $Q$ in
the RGE. Then one can derive the relation
\begin{equation}
\alpha_{S} = (b_{0}t)^{-1},~~~
t \equiv 2\ln\left[\left(\frac{\textstyle \langle
b_{k}\rangle}{\textstyle 2} + 1\right)\frac{\textstyle \langle
n_{\mbox{\scriptsize{J}}}\rangle m_{h}}{\textstyle
\Lambda}\right], \label{eq:My-new}
\end{equation}
if $\langle M_{\mbox{\scriptsize{J}}}\rangle$ is chosen as $Q$ for
some reaction at fixed $\sqrt{s}$, where $b_{0}=(33-2n_{f})/(12\pi)$ is the one-loop $\beta$-function coefficient, $n_{f}$ is number of quark flavors considered as light, $\Lambda$ is the QCD
parameter. That is why the distributions on $b_{k}$ are studied
usually. Thus the approach under discussion allows to distinguish,
at least, at qualitative level different dynamic mechanisms of
soft jet production in range of intermediate $\sqrt{s}$. The
separation of various dynamics of jet production is a difficult
task especially in non-perturbative region of $\sqrt{s}$. The
study of traditional collective observables in non-perturbative
initial energy domain allows to conclude for event shape mostly
\cite{Okorokov-IJMPA-27-1250037-2012}. On the other hand
investigation of jet properties with the aid of relativistically
invariant variables like (\ref{eq:1}) as well as observables
related with $b_{k}$ and defined above allows to make additional
suggestions concerning the dynamic features of mechanism of these
jet productions. This is additional improvement of
relativistically invariant method which seems (very) important at
intermediate $\sqrt{s}$ namely. Therefore approaches for analysis
of jet production based on traditional collective characteristics
and on relativistically invariant parameters (\ref{eq:1}) --
(\ref{eq:3}) are supplemented each other at intermediate
$\sqrt{s}$.

Because of (\ref{eq:1}) and (\ref{eq:3}) are dimensionless
parameters it seems the use of corresponding non-dimension mean
temperature $\langle \tilde{T}_{k}\rangle \equiv \langle
T_{k}\rangle / \sqrt{s_{0}}$ is more convenient without loss of
generality, where $s_{0}=1$ GeV$^{2}$. Thus in the present study
the set of dimensionless collective observables $\mathcal{G}
\equiv \{\mathcal{G}_{i}\}_{i=1}^{3}=\{\langle b_{k}\rangle,
\langle \tilde{T}_{k}\rangle, D\}$ characterized the dynamics of
creation of final state and its geometry in the space of
four-velocities is considered.

\section{Results: energy dependence for jet characteristics}\label{sec:3}
\begin{figure*}
\includegraphics[width=15.5cm,height=17.0cm]{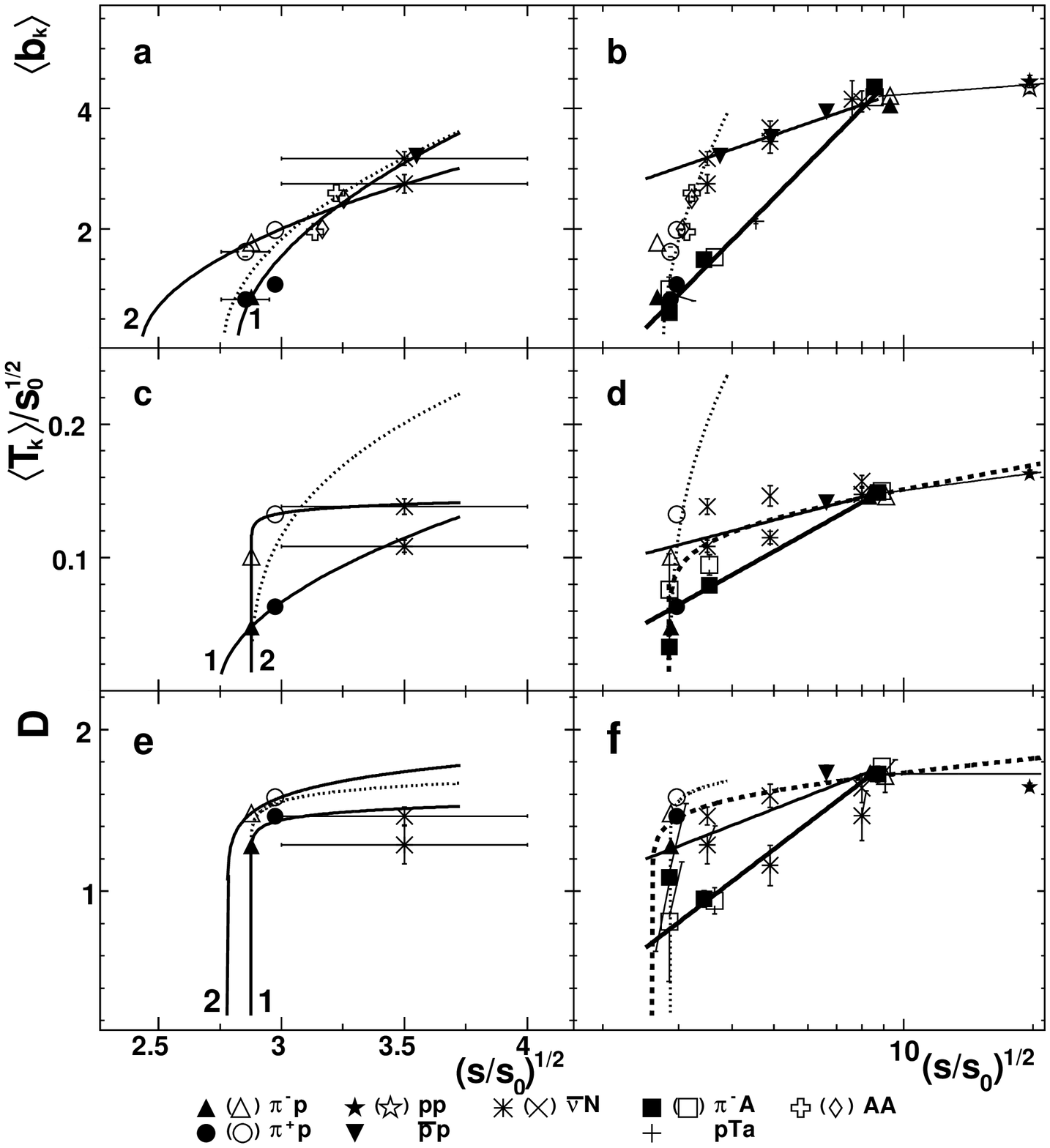}
\vspace*{8pt} \caption{Dependencies of parameters from the set
$\mathcal{G}$ on $\sqrt{s}$ at $\tilde{X}=0.1$ in the target
(beam) fragmentation region for various interactions. Experimental
data are from \cite{Okorokov-YaF-73-2016-2010} excepting values of
$\langle T_{k}\rangle$ and $D$ for $\bar{\nu}\mbox{N}$. The curve
1 (2) corresponds to the fit by power function of data samples for
target (beam) fragmentation region. Approximations for joint
samples for target and beam fragmentation regions are shown by
dotted line for experimental data on $\mbox{hh}$,
$\bar{\nu}\mbox{N}$ and $\mbox{AA}$ at $\sqrt{s} < 4$ GeV, by
solid line -- for $\mbox{hh}$ and $\bar{\nu}\mbox{N}$ reactions at
$4 \leq \sqrt{s} < 9$ GeV, by solid thick line -- for $\mbox{hA}$
collisions, by solid thin line -- for $\mbox{hh}$ and $\mbox{hA}$
interactions at $\sqrt{s} > 8$ GeV. Fits by function (\ref{eq:4})
of all available experimental data for $\langle T_{k}\rangle$ (d)
and $D$ (f) are shown by dashed lines.} \label{fig:1}
\end{figure*}

\begin{figure*}
\includegraphics[width=15.5cm,height=17.0cm]{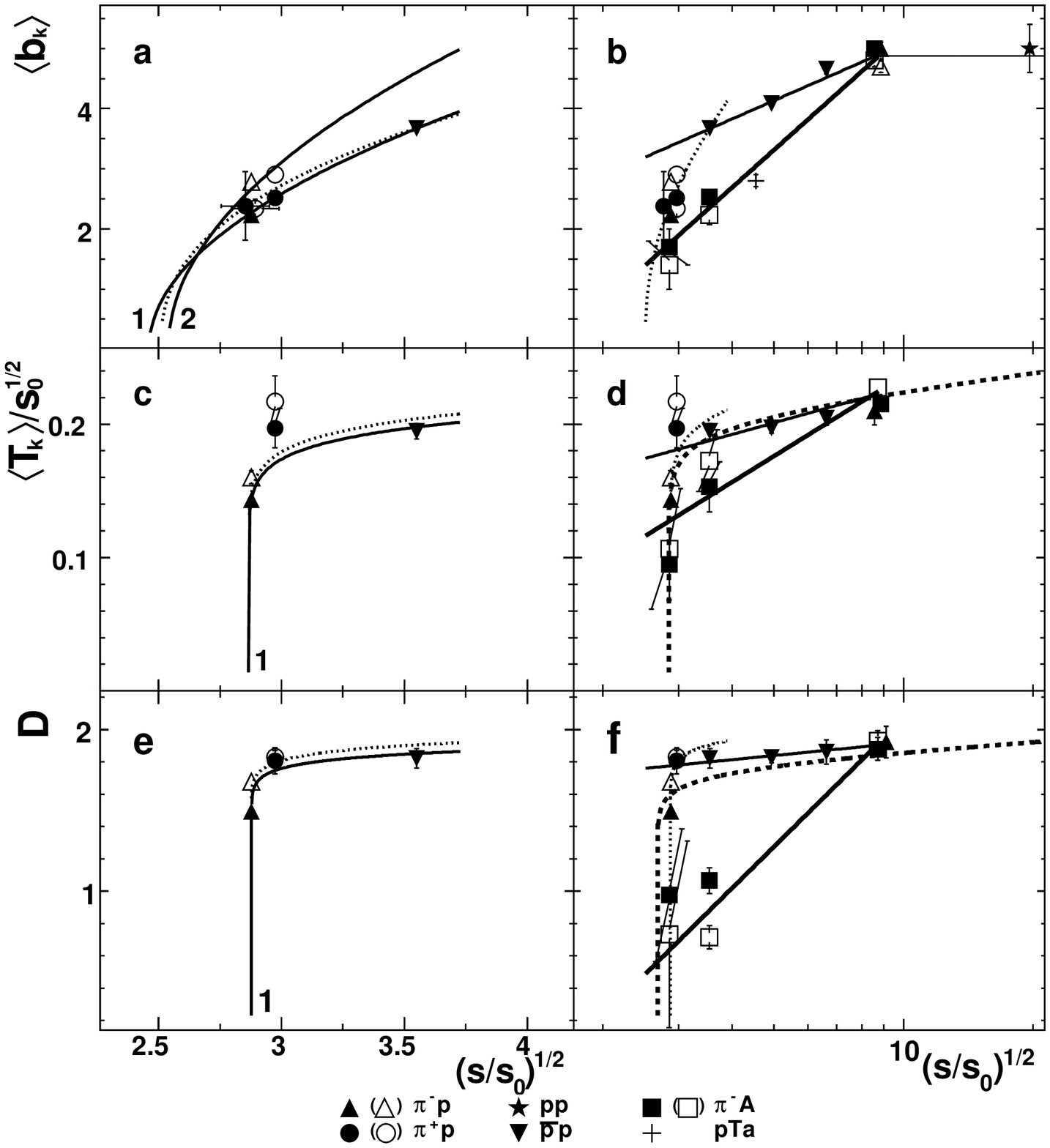}
\vspace*{8pt} \caption{Dependencies of parameters from the set
$\mathcal{G}$ on $\sqrt{s}$ at $\tilde{X}=0.2$ in the target
(beam) fragmentation region for various interactions. Experimental
data are from \cite{Okorokov-YaF-73-2016-2010} with exception of
the values of $\langle T_{k}\rangle$ and $D$ for
$\bar{\nu}\mbox{N}$. Notations for the experimental points and fit
curves are identical to that in Fig.\,\ref{fig:1}.} \label{fig:2}
\end{figure*}

Taking into account close values of $\langle \tilde{T}_{k}\rangle$
obtained for secondary $\pi^{-}$ and $\pi^{\pm}$ mesons in
\cite{Okorokov-YaF-62-1787-1999} as well as similar behavior of
distributions on $b_{k}$ observed for $\pi^{\pm}$ in $\bar{\nu}N$
interactions and for $\pi^{-}$ in the reactions with hadronic
beams at close initial energies
\cite{Baldin-FortschrPhys-38-1990-261}, in present study the
values of $\langle \tilde{T}_{k}\rangle$ and $D$ obtained for
$\pi^{\pm}$ in $\bar{\nu}N$ interactions are considered as
estimations of these parameters for $\pi^{-}$ mesons in the
corresponding reactions with $\bar{\nu}$ beam. This suggestion
allows to extend the experimental data base for $\langle
\tilde{T}_{k}\rangle$ and $D$ in comparison with
\cite{Okorokov-YaF-73-2016-2010}. One need to make the two
important comments. First of all, the estimations of $\langle
\tilde{T}_{k}\rangle$ and $D$ for $\bar{\nu}N$ interactions agree
reasonably both with results of another experiments at close
energies and with general trends (see Figs. \ref{fig:1},
\ref{fig:2} and detail discussion below). This is additional
evidence of applicability of the approach used for estimations of
all parameters from $\mathcal{G}$ in the case of $\bar{\nu}N$.
Second, cluster dimensions for $\bar{\nu}N$ as well as for large
number of reactions considered previously in
\cite{Okorokov-YaF-73-2016-2010} have fractional values for any
fragmentation regions. This observation confirms the conclusion
from \cite{Okorokov-YaF-73-2016-2010} and gives additional grounds
to suggest that soft pion jets demonstrate fractal-like properties
in wide class of interactions at intermediate energies $\sqrt{s} <
20$ GeV.

In the present study the dependence of all parameters from
$\mathcal{G}$ on initial energy has is investigated in detail
based on the available experimental data. In
\cite{Okorokov-YaF-62-1787-1999} the visible changing of behavior
of dependence $\langle b_{k}\rangle$ on $\sqrt{s}$ was observed.
Taking into account this feature and physical meaning of the
parameter (\ref{eq:1}) the hypothesis about changing of dynamic
regimes in processes of soft hadronic jet production at $\sqrt{s}
< 3 - 4$ GeV was suggested in \cite{Okorokov-YaF-62-1787-1999} and
confirmed some later in \cite{Okorokov-YaF-73-2016-2010}. In Fig.
\ref{fig:1} and Fig. \ref{fig:2} the dependencies of parameters
from $\mathcal{G}$ on $\sqrt{s}$ are shown for various
interactions at boundary values $\tilde{X}=0.1$ and 0.2
respectively. The data samples from
\cite{Okorokov-YaF-73-2016-2010} and estimations for parameters
$\langle T_{k}\rangle$, $D$ in the case of $\bar{\nu}N$ derived on
basis of differential distributions $dN/db_{k}$ from
\cite{Baldin-FortschrPhys-38-1990-261} are used. In the last case
the mean energy of hadronic final state, $W$, is considered as a
initial energy of reaction. Based on the shape of dependencies
under discussion and preceding studies
\cite{Okorokov-YaF-57-2225-1994,Okorokov-YaF-62-1787-1999,Okorokov-YaF-73-2016-2010}
various samples of experimental data are fitted by the power
function
\begin{equation}
\forall\, i:
~\mathcal{G}_{i}=a_{1}\bigl(\sqrt{s/s_{0}}-a_{2}\bigr)^{a_{3}},~~\sqrt{s/s_{0}}
\geq a_{2}\label{eq:4}
\end{equation}
in energy domain $\sqrt{s} < 4$ GeV for hadron-hadron and
nucleus-nucleus interactions and by the logarithmic function
\begin{equation}
\hspace{-2.5cm}\forall\, i:
~\mathcal{G}_{i}=a_{1}+a_{2}\ln(s/s_{0}). \label{eq:5}
\end{equation}
at higher energies for hadronic and $\bar{\nu}$ beams as well as
for hadron-nucleus collisions at any initial energies. One can
note that the shapes of energy dependencies for all collective
characteristics from $\mathcal{G}$ are similar at qualitative
level for fixed value of $\tilde{X}$.

\begin{table}[!b]
\caption{Fit results for $\mathcal{G}_{i}$ parameters for jointed
experimental data samples} \label{table:1}
\begin{center}
\begin{tabular}{|l|c|c|c|c|c|c|p{2.0cm}|} \hline
\multicolumn{1}{|c|}{Fit} &\multicolumn{3}{|c|}{$\tilde{X}=0.1$}
&\multicolumn{3}{|c|}{$\tilde{X}=0.2$}
\rule{0pt}{10pt}\\\cline{2-7} parameter & $\langle b_{k}\rangle$ &
$\langle \tilde{T}_{k}\rangle$ & $D$ & $\langle b_{k}\rangle$ &
$\langle \tilde{T}_{k}\rangle$ &
 $D$ \rule{0pt}{10pt}\\
\hline \multicolumn{7}{|c|}{$\mbox{hh},~\mbox{hA}$ interactions
($\sqrt{s} \! > \! 8$ GeV) \rule{0pt}{10pt}} \\
\hline
$a_{1}$        &$3.7 \pm 0.2$   & $0.108 \pm 0.008$ & $1.73 \pm 0.02$ & $4.88 \pm 0.05$ & -- & -- \rule{0pt}{10pt}\\
$a_{2}$        &$0.12 \pm 0.05$ & $0.009 \pm 0.002$ & $0.0$ (fixed)   & $0.0$ (fixed)   & -- & -- \\
$\chi^{2}$/NDF & 7.20           & 1.14              & 0.51            & 1.71            & -- & -- \\
\hline \multicolumn{7}{|c|}{$\mbox{hh},~\bar{\nu}\mbox{N}$
reactions
($4 \! \leq \! \sqrt{s} \! < \! 9$ GeV) \rule{0pt}{10pt}}\\
\hline
$a_{1}$        &$1.85 \pm 0.04$ & $0.070 \pm 0.016$ & $0.8 \pm 0.2$   & $1.94 \pm 0.08$ & $0.139 \pm 0.012$ & $1.68 \pm 0.13$ \rule{0pt}{10pt}\\
$a_{2}$        &$0.53 \pm 0.01$ & $0.018 \pm 0.004$ & $0.23 \pm 0.06$ & $0.68 \pm 0.03$ & $0.019 \pm 0.003$ & $0.06 \pm 0.04$ \\
$\chi^{2}$/NDF & 3.00           & 1.39              & 1.65            & 2.18            & 2.52              & 0.13 \\
\hline \multicolumn{7}{|c|}{$\mbox{hA}$ collisions
($\sqrt{s} \! < \! 9$ GeV) \rule{0pt}{10pt}} \\
\hline
$a_{1}$        &$-2.56 \pm 0.08$ & $-0.021 \pm 0.007$ & $-0.17 \pm 0.11$ & $-1.17 \pm 0.17$ & $0.04 \pm 0.03$   & $-0.56 \pm 0.14$ \rule{0pt}{10pt}\\
$a_{2}$        &$1.58 \pm 0.02$  & $0.039 \pm 0.002$  & $0.44 \pm 0.03$  & $1.39 \pm 0.05$  & $0.043 \pm 0.006$ & $0.57 \pm 0.04$ \\
$\chi^{2}$/NDF & 4.30            & 3.17               & 0.33             & 2.68             & 1.76              & 2.33 \\
\hline \multicolumn{7}{|c|}{$\mbox{hh},~\mbox{AA}$ interactions
($\sqrt{s} \!
< \! 4$ GeV) \rule{0pt}{10pt}} \\
\hline
$a_{1}$        &$3.69 \pm 0.02$ & $0.236 \pm 0.015$ & $1.68 \pm 0.05$   & $3.63 \pm 0.05$ & $0.210 \pm 0.006$ & $1.93 \pm 0.05$ \rule{0pt}{10pt}\\
$a_{2}$        &$2.76 \pm 0.01$ & $2.877 \pm 0.001$ & $2.875 \pm 0.001$ & $2.51 \pm 0.03$ & $2.865 \pm 0.007$ & $2.875 \pm 0.001$ \\
$a_{3}$        &$0.49 \pm 0.01$ & $0.34$ (fixed)    & $0.04 \pm 0.01$   & $0.40 \pm 0.01$ & $0.080 \pm 0.004$ & $0.031 \pm 0.007$ \\
$\chi^{2}$/NDF & 133            & 8.42              & 8.21              & 9.25            & 7.17              & 2.84 \\
\hline
\end{tabular}
\end{center}
\end{table}

\begin{table}[!h]
\caption{Fit results for $\mathcal{G}_{i}$ in $\sqrt{s} \! < \! 4$
GeV energy domain for separated experimental data samples}
\label{table:2}
\begin{center}
\begin{tabular}{|l|c|c|c|c|c|c|p{2.0cm}|} \hline
\multicolumn{1}{|c|}{Fit} &\multicolumn{3}{|c|}{$\tilde{X}=0.1$}
&\multicolumn{3}{|c|}{$\tilde{X}=0.2$}
\rule{0pt}{10pt}\\\cline{2-7} parameter & $\langle b_{k}\rangle$ &
$\langle \tilde{T}_{k}\rangle$ & $D$ & $\langle b_{k}\rangle$ &
$\langle \tilde{T}_{k}\rangle$ &
 $D$ \rule{0pt}{10pt}\\
\hline \multicolumn{7}{|c|}{target fragmentation region \rule{0pt}{10pt}} \\
\hline
$a_{1}$        &$3.79 \pm 0.03$ & $0.132 \pm 0.019$& $1.53 \pm 0.05$   & $3.52 \pm 0.07$ & $0.204 \pm 0.006$ & $1.87 \pm 0.04$ \rule{0pt}{10pt}\\
$a_{2}$        &$2.82 \pm 0.02$ & $2.75 \pm 0.04$  & $2.875 \pm 0.004$ & $2.46 \pm 0.04$ & $2.865 \pm 0.007$ & $2.877 \pm 0.001$ \\
$a_{3}$        &$0.52 \pm 0.05$ & $0.5$ (fixed)    & $0.03$ (fixed)    & $0.5$ (fixed)   & $0.08$ (fixed)    & $0.03$ (fixed)\\
$\chi^{2}$/NDF & 5.70              & 0.03             & 3.90              & 0.09            & 0.06              & 0.80 \\
\hline
\multicolumn{7}{|c|}{beam fragmentation region \rule{0pt}{10pt}} \\
\hline
$a_{1}$        &$2.65 \pm 0.13$ & $0.142 \pm 0.003$ & $1.79 \pm 0.05$ & $5 \pm 2$     & -- & -- \rule{0pt}{10pt}\\
$a_{2}$        &$2.43 \pm 0.04$ & $2.877 \pm 0.001$ & $2.78 \pm 0.04$ & $2.5 \pm 0.3$ & -- & -- \\
$a_{3}$        &$0.5 \pm 0.4$   & $0.03$ (fixed)    & $0.08$ (fixed)  & $0.5$ (fixed) & -- & -- \\
$\chi^{2}$/NDF & 5.19           & 0.06              & 6.43            & 2.29          & -- & -- \\
\hline
\end{tabular}
\end{center}
\end{table}

For present study the domain $\sqrt{s} < 4$ GeV is important
especially. Therefore results of experiments with different beam
types with exception of the hadron-nucleus collisions and
corresponding approximations are shown for range $\sqrt{s} < 4$
GeV separately in Figs. \ref{fig:1}a, c, e for boundary value
$\tilde{X}=0.1$ and in Figs. \ref{fig:2}a, c, e at
$\tilde{X}=0.2$. The uncertainties of initial momenta are taking
into account in the pictures indicated above. At soft cut
$\tilde{X}=0.1$ values of $\langle b_{k}\rangle$ parameter for
symmetric nucleus-nucleus collisions are in a good agreement with
the general trend (Figs. \ref{fig:1}a, b). At initial energies
$\sqrt{s} \simeq 3$ GeV and boundary cut $\tilde{X}=0.1$ values of
$\langle b_{k}\rangle$ (Figs. \ref{fig:1}a, b) and $\langle
\tilde{T}_{k}\rangle$ (Figs. \ref{fig:1}c, d) are smaller
significantly than that in another interactions at slightly larger
energies. The behavior of $D(\sqrt{s})$ at these energies does not
contradict with that conclusion. At harder cut value of
$\tilde{X}$ for distinguishing of fragmentation regions all
parameters from $\mathcal{G}$ show similar behavior at $\sqrt{s}
\simeq 3$ GeV. Thus $\forall~i=1-3$ dependencies
$\mathcal{G}_{i}(\sqrt{s})$ demonstrate the visible changing of
its shapes in narrow range $\sqrt{s} \simeq 3$ GeV both at
$\tilde{X}=0.1$ (Fig. \ref{fig:1}) and at $\tilde{X}=0.2$ (Fig.
\ref{fig:2}). As indicated above the experimental dependencies
$\mathcal{G}_{i}(\sqrt{s})$, $i=1-3$ are fitted by power function
(\ref{eq:4}) at initial energies $\sqrt{s} < 4$ GeV. Fit results
for samples jointed for different fragmentation regions are shown
in Fig. \ref{fig:1} and Fig. \ref{fig:2} by dotted lines, the
numerical values of fit parameters are presented in Table
\ref{table:1}. Results for $\langle b_{k}\rangle$ are from
\cite{Okorokov-YaF-62-1787-1999,Okorokov-YaF-73-2016-2010}. In
\cite{Okorokov-YaF-62-1787-1999,Okorokov-YaF-73-2016-2010,Okorokov-ISHEPP-154-2006}
was shown that properties of soft pion jets are depended on
fragmentation region in domain $\sqrt{s}$ under consideration.
Such dependence results in significant difference of $\langle
b_{k}\rangle$ values for various fragmentation regions in certain
reaction. The analogous situation is observed for other parameters
from $\mathcal{G}$ here. Therefore the sharp behavior of energy
dependencies and spread of experimental points give no way to
attain statistically acceptable quality of fits (as usual,
$\chi^{2}/\mbox{NDF} \sim 10$). However, as seen from Fig.
\ref{fig:1} and Fig. \ref{fig:2} the function (\ref{eq:4}) is in
rather good agreement, at leats, at qualitative level with
experimental data at all $\tilde{X}$ values considered above. It
is important to note that, in general case, a power-law behavior
is peculiar to the transition region, where new degrees of freedom
of the system under study come into relevant and play a role
amplified successively \cite{Okorokov-YaF-73-2016-2010}. The
parameter $a_{2}$ in (\ref{eq:4}) can be put in correspondence to
the energy (in GeV), $\sqrt{s_{c}}$, at which the internal
structure of interacting hadrons, i.e. quark-gluon degrees of
freedom, begins to manifest itself experimentally. The values of
dimensionless parameter $a_{2}$, i.e. values of low boundary
energy $\sqrt{s_{c}}$ in GeV for color degrees of freedom
experimental appearance in jet production, are shown in Table
\ref{table:1} for all members of set $\mathcal{G}$ at various
$\tilde{X}$. One can see that the values of $a_{2}$ derived for
$\langle \tilde{T}_{k}\rangle$ and $D$ agree with each other and
are some larger than values of the fit parameter obtained for
energy dependence of $\langle b_{k}\rangle$ at fixed $\tilde{X}$.
The such disagreement can be caused by some physical reasons which
can result in more sharp increasing of $\langle
\tilde{T}_{k}\rangle$ and $D$ in comparison with $\langle
b_{k}\rangle$ as well as some methodological features, for
example, smaller volumes of experimental samples in the case of
former two parameters. Additional new experimental data in initial
energy domain $\sqrt{s} \leq 4$ ÃýÂ seems important for detail
investigation of behavior of dependencies
$\mathcal{G}_{i}(\sqrt{s})$, $i=1-3$ and for decreasing of
uncertainties in physical conclusions. In the energy range under
consideration available data samples were fitted by function
(\ref{eq:4}) for the region of target fragmentation (curves 1 in
Figs. \ref{fig:1}a, c, e and in Figs. \ref{fig:2}a, c, e) and beam
fragmentation (curves 2 in Figs. \ref{fig:1}a, c, e and in Fig.
\ref{fig:2}a) individually for various values of cut $\tilde{X}$.
The numerical values of fit parameters are shown in Table
\ref{table:2}. Approximations for parameters $\langle
\tilde{T}\rangle$ and $D$ in the case of beam fragmentation at
$\tilde{X}=0.2$ are impossible because of absent of necessary
amount of data points (Figs. \ref{fig:2}c, e). As seen there is
significant improvement of fit qualities for any fragmentation
region and value of $\tilde{X}$ ($\chi^{2}/\mbox{NDF} \sim 3-5$),
moreover statistically acceptable values of $\chi^{2}/\mbox{NDF}$
are succeeded for 50\% of samples under study. Taking into account
relatively small volumes of available experimental data samples
for $\langle \tilde{T}_{k}\rangle$ and $D$, especially at
$\tilde{X}=0.2$, and shapes of $\langle
\tilde{T}_{k}\rangle(\sqrt{s})$ and $D(\sqrt{s})$ the energy
dependencies for these collective parameters were fitted by
function (\ref{eq:4}) in the full energy domain $\sqrt{s} > 2$
GeV. Obviously the sample volumes increase significantly for such
energy domain with respect to the samples fitted for separate
energy ranges with results shown in Tables \ref{table:1},
\ref{table:2}. The approximations for $\sqrt{s} > 2$ GeV were made
for samples jointed by fragmentation regions as well as for target
and beam fragmentation regions separately. The fits of $\langle
\tilde{T}_{k}\rangle(\sqrt{s})$ in the first case are shown by
dashed lines in Fig. \ref{fig:1}d at $\tilde{X}=0.1$ and for
harder kinematic cut -- in Fig. \ref{fig:2}d; numerical values of
fit parameters at $\tilde{X}=0.1 (0.2)$ are following:
$a_{1}=0.1168 \pm 0.0005 (0.1946 \pm 0.0015)$, $a_{2}=2.854 \pm
0.001 (2.854 \pm 0.001)$, $a_{3}=0.131 \pm 0.002 (0.071 \pm
0.004)$, $\chi^{2} / \mbox{NDF}=266 (3.50)$. As seen the only
qualitative agreement is observed between fit function
(\ref{eq:4}) and experimental data at $\tilde{X}=0.1$ (Fig.
\ref{fig:1}d) so long as fit quality is significantly poorer than
that at $\tilde{X}=0.2$ (Fig .\ref{fig:2}d). The fits of
$D(\sqrt{s})$ for the jointed samples are shown by dashed lines in
Figs. \ref{fig:1}f, \ref{fig:2}f at $\tilde{X}=0.1$ and $0.2$
respectively; numerical values of fit parameters at $\tilde{X}=0.1
(0.2)$ are following: $a_{1}=1.539 \pm 0.007 (1.71 \pm 0.02)$,
$a_{2}=2.60 \pm 0.03 (2.68 \pm 0.03)$, $a_{3}=0.059 \pm 0.003
(0.041 \pm 0.005)$, $\chi^{2} / \mbox{NDF}=28 (32)$. For this
collective parameter the fit qualities allow to consider
qualitative agreement only between function (\ref{eq:4}) and
experimental data (Figs. \ref{fig:1}f, \ref{fig:2}f). Separate
approximations by (\ref{eq:4}) of $\langle
\tilde{T}_{k}\rangle(\sqrt{s})$ result in the following results at
$\tilde{X}=0.1 (0.2)$ for target fragmentation $a_{1}=0.1014 \pm
0.0007 (0.189 \pm 0.002)$, $a_{2}=2.853 \pm 0.001 (2.854 \pm
0.001)$, $a_{3}=0.200 \pm 0.003 (0.070 \pm 0.005)$, $\chi^{2} /
\mbox{NDF}=10 (2.04)$ and for beam fragmentation $a_{1}=0.1403 \pm
0.0012 (0.204 \pm 0.003)$, $a_{2}=2.877 \pm 0.001 (2.854 \pm
0.001)$, $a_{3}=0.033 \pm 0.005 (0.062 \pm 0.006)$, $\chi^{2} /
\mbox{NDF}=8.36 (2.71)$. As expected the fit qualities improve
significantly with respect to the fits of jointed samples, the
values of $\sqrt{s_{c}}$ are equal within errors for various
approximation approaches with exception of beam fragmentation at
$\tilde{X}=0.1$. The shape of $D(\sqrt{s})$ (Figs. \ref{fig:1}f,
\ref{fig:2}f) and detail study show that the presence of
experimental data for $\pi^{-}\mbox{Ne}$ interactions at initial
momentum $p_{0}=6.2$ GeV/$c$
\cite{Okorokov-YaF-57-2225-1994,Okorokov-YaF-73-2016-2010} in
sample under fit leads to significant poor fit quality and to weak
changing of $\sqrt{s_{c}}$ value at the same time. Therefore the
separate fits of $D(\sqrt{s})$ in full energy domain were made for
samples with excepted of experimental points for
$\pi^{-}\mbox{Ne}$ reaction. The following results are obtained at
$\tilde{X}=0.1 (0.2)$ for target fragmentation $a_{1}=1.559 \pm
0.014 (1.78 \pm 0.02)$, $a_{2}=2.854 \pm 0.004 (2.858 \pm 0.001)$,
$a_{3}=0.051 \pm 0.003 (0.043 \pm 0.004)$, $\chi^{2} /
\mbox{NDF}=3.98 (1.44)$ and for beam fragmentation $a_{1}=1.658
\pm 0.001 (1.89 \pm 0.04)$, $a_{2}=2.9 \pm 0.4 (2.874 \pm 0.001)$,
$a_{3}=0.029 \pm 0.002 (0.013 \pm 0.016)$, $\chi^{2} /
\mbox{NDF}=3.08 (1.78)$. As seen the fit qualities are improved
significantly, moreover statistically acceptable values of
$\chi^{2}/\mbox{NDF}$ are succeeded for samples under study at
$\tilde{X}=0.2$. Therefore the approximations of experimental
samples for $\langle \tilde{T}_{k}\rangle$ and $D$ in full energy
domain allows to get, in particular, the estimations for $a_{2}$
for any fragmentation regions and values of $\tilde{X}$ considered
here.

The values of $\sqrt{s_{c}}$ corresponded to the approximations of
energy dependence of parameters from $\mathcal{G}$ with accounts
of optimal combinations of fit qualities and volumes of available
experimental data samples are summarized in the Table
\ref{table:3}.

\begin{table}[!h]
\caption{The values of low boundary energy $\sqrt{s_{c}}$ in GeV}
\label{table:3}
\begin{center}
\begin{tabular}{|l|c|c|c|c|p{2.0cm}|} \hline
\multicolumn{1}{|c|}{Parameter} &\multicolumn{2}{|c|}{target
fragmentation, $\tilde{X}$} &\multicolumn{2}{|c|}{beam
fragmentation, $\tilde{X}$} \rule{0pt}{10pt}\\\cline{2-5} from
$\mathcal{G}$ & 0.1 & 0.2 & 0.1
& 0.2 \rule{0pt}{10pt}\\
\hline
$\langle b_{k}\rangle$         &$2.82 \pm 0.02$   & $2.46 \pm 0.04$   & $2.43 \pm 0.04$   & $2.5 \pm 0.3$     \rule{0pt}{10pt}\\
$\langle \tilde{T}_{k}\rangle$ &$2.853 \pm 0.001$ & $2.854 \pm 0.001$ & $2.877 \pm 0.001$ & $2.854 \pm 0.001$ \\
$D$                            &$2.854 \pm 0.002$ & $2.858 \pm 0.001$ & $2.9 \pm 0.4$     & $2.874 \pm 0.001$ \\
\hline
\end{tabular}
\end{center}
\end{table}

As seen from Table \ref{table:3} the values of $\sqrt{s_{c}}$
agree reasonably for different kinematic cuts ($\tilde{X}$) and
fragmentation regions. It should be emphasized that the results
obtained by study of fractal properties of pion jets are in a good
agreement with results of physical analysis of energy dependencies
of parameters $\langle b_{k}\rangle$ and $\langle
\tilde{T}_{k}\rangle$. Thus this study extended on the full
parameter set $\mathcal{G}$ increases the degree of confidence of
results obtained earlier
\cite{Okorokov-YaF-62-1787-1999,Okorokov-YaF-73-2016-2010} and
gives the additional evidences in favor of the suggested
hypothesis concerned the changing of dynamic regimes in
multiparticle production processes caused by the onset of
experimental manifestation of quark degrees of freedom in soft jet
production at $\sqrt{s} \sim 3$ GeV and the corresponding
transition from description of strong interaction with the aid of
meson-baryon degrees of freedom to the using of quark-gluon ones.
For the first time the quantitative estimation is obtained for the
low boundary on energy at which the quark degrees of freedom start
to manifest itself experimentally in production of soft pion jets.
The value of estimation of parameter under discussion matched with
taking into account of results for all collective characteristics
from $\mathcal{G}$ is $\sqrt{s_{c}} = (2.8 \pm 0.6)$ GeV. In
\cite{Okorokov-YaF-76-2013} the following qualitative estimation
was obtained for universal low boundary of experimental
manifestation of jet shape of final state in multiparticle
production processes: $\sqrt{s_{\mbox{\scriptsize{min}}}} \sim 3$
GeV. This qualitative estimation is equal to $\sqrt{s_{c}}$ within
errors. It should be noted that in despite of possibility for
production of localized and separated groups of secondary
particles in the framework of models corresponded to meson-baryon
level of matter the conception of "hadron jet" itself is deeply
relates with QCD because the jet is usually defined as group of
particles produced due to hadronization of common color charge.
Therefore it can not be excluded the possibility that the
parameters $\sqrt{s_{\scriptsize{\mbox{min}}}}$ and $\sqrt{s_{c}}$
characterize the united physical effect -- the onset of
experimental manifestation of quark degrees of freedom in soft
processes of multiparticle production and, as consequence, jet
event shape. Therefore the results obtained by using of
traditional and relativistically invariant collective observables
agree and add each other.

Earlier in \cite{Okorokov-YaF-62-1787-1999}, a statistically
reasonable description by function (\ref{eq:5}) was obtained for
experimental dependencies $\langle b_{k}\rangle(\sqrt{s})$ in the
case of joined samples for hadron-hadron and $\bar{\nu}\mbox{N}$
reactions (solid lines in Figs. \ref{fig:1}b, \ref{fig:2}b), as
well as for hadron-nucleus interactions (solid thick lines in
Figs. \ref{fig:1}b, \ref{fig:2}b) in the range $3.5 \! < \!
\sqrt{s} < 9$ GeV; for hadron-hadron and hadron-nucleus collisions
at higher energies (solid thin lines in Figs. \ref{fig:1}b,
\ref{fig:2}b). Additionally in the present study it is obtained
that quality of approximation of dependence $\langle
b_{k}\rangle(\sqrt{s})$ by generalized logarithmic function
$a_{1}+a_{2}[\ln(s/s_{0})]^{a_{3}}$ improves at fixed value
$a_{3}=1.0$ for fitting procedure in both energy domains $3.5 <
\sqrt{s} < 9$ GeV and $\sqrt{s} > 8$ GeV. Thus, the dependence
$\langle b_{k}\rangle(\sqrt{s})$ admits a universal approximation
by (\ref{eq:5}) in domain $\sqrt{s} > 3.5$ GeV for a wide class of
interactions at any $\tilde{X}$ values studied here. The results
of quantitative analysis of $\langle b_{k}\rangle$ agree with
qualitative hypothesis about slow (logarithmic) growth of jet size
and estimation of $\langle b_{k}\rangle$ at $\sqrt{s} \geq 10$ GeV
indicated in \cite{Grishin-Book-1988}. Taking into account
qualitative shapes of dependencies $\langle
\tilde{T}_{k}\rangle(\sqrt{s})$ and $D(\sqrt{s})$ in energy ranges
under consideration the experimental samples joined on
fragmentation regions for these members of $\mathcal{G}$ were
fitted by function (\ref{eq:5}). The fit results are presented in
Table \ref{table:1}. As seen both $\langle b_{k}\rangle$ and other
parameters from $\mathcal{G}$ increase faster for hadron-nucleus
reactions than that for nucleonic target in the range of initial
energies under study. Results for $\langle \tilde{T}_{k}\rangle$
and $D$ obtained for hadron-nuclear collisions at $\sqrt{s} \sim
3-5$ GeV can be considered as indication on sensitivity of nuclear
target on dynamic (mean temperature) and geometric (cluster
dimension) parameters of hadronic system passes through nucleus.
This observation allows to generalize the conclusion about
influence of nuclear matter on size of soft pion jets
\cite{Okorokov-YaF-57-2225-1994} on full set of collective
parameters $\mathcal{G}$. This assumption is in good agreement
with estimations both for fragmentation length at $\sqrt{s} \sim
3-5$ GeV \cite{Okorokov-YaF-57-2225-1994} and for color field
formation length
$L^{\mbox{\footnotesize{c.f.}}}_{\mbox{\footnotesize{f}}} \sim
Rb_{k}\,/2$, where $R$ is the radius of target nucleus
\cite{Baldin-FortschrPhys-38-1990-261}. In particular, the last
relation indicates that
$L^{\mbox{\footnotesize{c.f.}}}_{\mbox{\footnotesize{f}}} \lesssim
R$ and nucleus can influence on jet properties at values $b_{k}
\lesssim 2$ which correspond the hadron-nuclear reactions in
energy domain $\sqrt{s} \sim 3-5$ GeV (Figs. \ref{fig:1}b,
\ref{fig:2}b).

In initial energy domain $\sqrt{s} > 8$ GeV the samples of
available experimental data for $\mathcal{G}_{i}$, $i=1-3$ have a
significantly smaller volumes than that at $\sqrt{s} \leq 8$ GeV.
Therefore the experimental points at $p_{0}$=40 GeV/$c$ are used
as boundary for "linkage" of fit results both for energies
$\sqrt{s} > 8$ GeV and for lower energies. Accounting for
suggestions about universal jet properties in hadron-hadron and
hadron-nucleus interactions \cite{Baldin-FortschrPhys-38-1990-261}
the experimental data both for $\pi^{-}\mbox{p}$ and for
$\pi^{-}\mbox{C}$ reaction at $P_{0}$=40 GeV/$c$ are included in
the sample under fit. In consequence of absent of experimental
data the approximations of $\langle
\tilde{T}_{k}\rangle(\sqrt{s})$ and $D(\sqrt{s})$ in separate
energy domain $\sqrt{s}
> 8$ GeV are possible at soft cut for $\tilde{X}$ only
(Figs. \ref{fig:1}d, f). Fit results are shown in Table
\ref{table:1}. The value of $a_{2}$ parameter in (\ref{eq:5}) is
equal to zero within statistical errors for fit of $D(\sqrt{s})$
in energy domain under study. Therefore the experimental sample is
fitted at fixed value $a_{2}=0$ and results are shown in Table
\ref{table:1}. On the other hand the slow logarithmic growth for
$D$ in accordance with (\ref{eq:5}) can not be excluded
unambiguously because of small volume of sample of available
experimental data.

Therefore $\forall~i=1-3$ dependencies $\mathcal{G}_{i}(\sqrt{s})$
admit a universal approximation in the form (\ref{eq:5}) at
$\sqrt{s} > 3.5$ GeV for a wide class of interactions at any
$\tilde{X}$ values considered here.

The comparative analysis collective properties ($\langle
b_{k}\rangle$, $\langle T_{k}\rangle$) of soft pion jets and
proton clusters
\cite{Baldin-Lektcii-43-P1-87-912-1987,Baldin-FortschrPhys-38-1990-261,Baldin-ECHAYA-29-577-1998}
at intermediate energies allows to get the following conclusions.
The proton clusters in hadron-nucleus and nucleus-nucleus
collisions are characterized significantly smaller values of
$\langle b_{k}\rangle$ than pion jets. The mean temperature of
particles in clusters is $\langle T_{k}\rangle \sim 65$ MeV for
target fragmentation in $p\mbox{C}$, $d\mbox{C}$ and $\mbox{CC}$
(clusters of the first type) collisions at $p_{0}$=4.2\,A GeV/$c$
\cite{Baldin-Lektcii-43-P1-87-912-1987,Baldin-FortschrPhys-38-1990-261}
which is comparable with $\langle T_{k}\rangle$ of pion jets in
this fragmentation region for $\pi^{+}p$ reactions at the
correspondent initial momentum $p_{0}$=4.2 GeV/$c$
\cite{Okorokov-YaF-62-1787-1999}. The temperature of second type
clusters produced in $\mbox{CC}$ collisions due to multi-nucleon
interactions in comparison with $\pi^{+}p$ interactions
\cite{Okorokov-YaF-62-1787-1999} is significantly larger than that
for pion jets in target fragmentation region and is equal to the
$\langle T_{k}\rangle$ of pion jets for beam fragmentation within
errors. These relations between mean temperatures of pion jets and
proton clusters can be dominated, in particular, significantly
larger proton mass ($m_{p}$) that pion one and definition
(\ref{eq:2}) re-written for proton clusters
\cite{Baldin-Lektcii-43-P1-87-912-1987,Baldin-FortschrPhys-38-1990-261}.

\section{Summary}\label{sec:4}
The following conclusions can be obtained by summarizing of the
basic results of the present study.

The qualitative relation between jet size in the space of relative
four-velocities and strong coupling constant is derived in the
lowest order of RGE.

The energy dependencies for all parameters from set $\mathcal{G}$
show the similar behaviors at qualitative level. These
dependencies are described by power function at $\sqrt{s} \leq 4$
GeV and logarithmic function at $\sqrt{s} > 3.5$ GeV reasonably.

The behavior of dependencies of $\langle b_{k}\rangle$, $\langle
T_{k}\rangle$ and $D$ on initial energy at $\sqrt{s} \sim 3$ GeV,
possibly, is dominated by the onset of experimental manifestation
of quark degrees of freedom in production of soft pion jets and
corresponding transition from the description of these processes
in terms of meson-baryon degrees of freedom to using of color
(quark-gluon) degrees of freedom. For the first time the
quantitative estimations are obtained for the low boundary energy
at which quark degrees of freedom start to display itself in
production of soft pion jets experimentally. The value of
estimation for this parameter, matched with taking into account of
all collective characteristics used in present study, is
$\sqrt{s_{c}} = (2.8 \pm 0.6)$ GeV.

The influence of presence of nuclear matter on dynamic and
geometric characteristics of soft pion jets in the space of
relative four-dimensional velocities is observed at $\sqrt{s} \sim
3-5$ GeV.

\section*{Acknowledgments}

The author is grateful to Prof. A.\,A. Petrukhin and Prof. A.\,K.
Ponosov for useful discussions.

\newpage
\newpage


\begin{thebibliography}{999}
\bibitem{Okorokov-IJMPA-27-1250037-2012}
V.\,A. Okorokov, Int. J. Mod. Phys. A {\bf 27}, 1250037 (2012).
\bibitem{Baldin-DoklANUSSR-222-1064-1975}
A.\,M. Baldin, Dokl. Akad. Nauk SSSR {\bf 222}, 1064 (1975) [Sov.
Phys. Dokl. {\bf 20}, 418 (1975)]; \emph{ibid} {\bf 279}, 1352
(1984) [\emph{ibid} {\bf 29}, 1031 (1984)].
\bibitem{Okorokov-PhD-1996}
V.\,A. Okorokov, \emph{Collective characteristics of multiparticle
production processes in hadron-hadron and hadron-nuclear interactions in domain of intermediate energies}.
PhD thesis, 
MEPhI, Moscow, 1996.
\bibitem{Okorokov-YaF-57-2225-1994}
I.\,L. Kiselevich, V.\,I. Mikhailichenko, V.\,A. Okorokov \emph{et
al.}, Yad. Fiz. {\bf 57}, 2225 (1994) [Phys. At. Nucl. {\bf 57},
2140 (1994)].
\bibitem{Okorokov-YaF-62-1787-1999}
V.\,I. Mikhailichenko, V.\,A. Okorokov, A.\,K. Ponosov, F.\,M.
Sergeev, Yad. Fiz. {\bf 62}, 1787 (1999) [Phys. At. Nucl. {\bf
62}, 1665 (1999)].
\bibitem{Baldin-YaF-44-1209-1986}
A.\,M. Baldin, B.\,V. Batyunya, I.\,M. Gramenetckii \emph{et al.},
Yad. Fiz. {\bf 44}, 1209 (1986) [Sov. J. Nucl. Phys. {\bf 44}, 785
(1987)].
\bibitem{Baldin-PreprintJINR-E1-87-142-1987}
A.\,M. Baldin, L.\,A. Didenko, V.\,G. Grishin \emph{et al.},
Preprint JINR E1-87-142, Dubna (1987).
\bibitem{Baldin-Lektcii-43-P1-87-912-1987}
A.\,M. Baldin, L.\,A. Didenko, \emph{Asymptotic properties of
hadronic matter in the space of four-dimensional relative
velocities}. Lectures for young scientists {\bf 43}, JINR, Dubna,
1987.
\bibitem{Baldin-FortschrPhys-38-1990-261}
A.\,M. Baldin, L.\,A.Didenko, Fortschr. Phys. {\bf 38}, 261
(1990).
\bibitem{Baldin-ECHAYA-29-577-1998}
A.\,A. Baldin, A.\,M. Baldin, Fiz. Elem. Chastits At. Yadra {\bf
29}, 577 (1998) [Phys. Part. Nucl. {\bf 29}, 232 (1998)].
\bibitem{Okorokov-NSMEPhI-218-2000}
V.\,A. Okorokov, A.\,K. Ponosov, F.\,M. Sergeev, in {\sl
Scientific session MEPhI-2000}, MEPhI, V.{\bf 7}, 2000, p.218.
\bibitem{Okorokov-YaF-73-2016-2010}
V.\,A. Okorokov, A.\,K. Ponosov, F.\,M. Sergeev, Yad. Fiz. {\bf
73}, 2016 (2010) [Phys. At. Nucl. {\bf 73}, 1963 (2011)].
\bibitem{Fomenko-book-1998}
A.\,T. Fomenko, \emph{Visual geometry and topology: mathematical
images in the real world}, Moscow, MSU, 1998.
\bibitem{Baldin-NPA-434-695c-1985}
A.\,M. Baldin, Nucl. Phys. A {\bf 434}, 695c (1985).
\bibitem{Grishin-Book-1988}
V.\,G. Grishin, \emph{Quarks and hadrons in high energy particle
interactions}, Moscow, Energoatomizdat, 1988.
\bibitem{Okorokov-ISHEPP-154-2006}
V.\,A. Okorokov, in {\sl XVIII International Baldin Seminar on
High Energy Physics Problems "Relativistic Nuclear Physics and
Quantum Chromodynamics"}, Eds. by A.\,N. Sissakian, V.\,V. Burov,
A.\,I. Malakhov, Dubna, V.{\bf I}, 2008, p.154.
\bibitem{D0-PLB-718-56-2012}
V.\,M. Abazov, B. Abbott, B.\,S. Acharya \emph{et al.}, Phys.
Lett. B {\bf 718}, 56 (2012).
\bibitem{CMS-arXiv-1304.7498-2013}
S. Chatrchyan, V. Khachatryan, A.\,M. Sirunyan \emph{et al.},
arXiv: 1304.7498 [hep-ex].
\bibitem{Okorokov-YaF-76-2013}
V.\,A. Okorokov, A.\,K. Ponosov, Yad. Fiz. {\bf 76}, 2013 (to be
published).

\end{thebibliography}
\end{document}